\documentclass[acmlarge, nonacm]{acmart}

\AtBeginDocument{%
  \providecommand\BibTeX{{%
    \normalfont B\kern-0.5em{\scshape i\kern-0.25em b}\kern-0.8em\TeX}}}

\setcopyright{acmcopyright}

\acmJournal{IMWUT}
\acmYear{0000}
\acmMonth{0}
\acmVolume{0}
\acmNumber{0}
\acmArticle{0}
\acmArticleSeq{00}
\acmDOI{10.1145/1122445.1122456}

\usepackage{xspace}
\usepackage{amsmath}
\usepackage{url}
\usepackage{arydshln}

\newcommand*{\networkname}{ATS-UNet\xspace}

\begin{document}

\title{Enabling Real-time On-chip Audio Super Resolution for Bone Conduction Microphones}
\author{Yuang Li}
\email{leo.liyuang@gmail.com}
\affiliation{%
  \institution{Beijing University of Posts and Telecommunications}
  \city{Haidian Qu}
  \state{Beijing Shi}
  \country{China}
}

\author{Yuntao Wang}
\email{yuntaowang@tsinghua.edu.cn}
\affiliation{%
  \institution{Tsinghua Univiersity}
  \city{Haidian Qu}
  \state{Beijing Shi}
  \country{China}
}

\author{Xin Liu}
\email{xliu0@cs.washington.edu}
\affiliation{%
  \institution{University of Washington}
  \city{Seattle}
  \state{WA}
  \country{USA}
}

\author{Yuanchun Shi}
\email{shiyc@tsinghua.edu.cn}
\affiliation{%
  \institution{Tsinghua Univiersity}
  \city{Haidian Qu}
  \state{Beijing Shi}
  \country{China}
}

\author{Shao-Fu Shih}
\email{shaofu.shih@hotmail.com}
\affiliation{%
  \institution{Harman International}
  \city{Stamford}
  \state{Connecticut}
  \country{USA}
}
\begin{abstract}
Voice communication using the air conduction microphone in noisy environments suffers from the degradation of speech audibility. Bone conduction microphones (BCM) are robust against ambient noises but suffer from limited effective bandwidth due to their sensing mechanism. Although existing audio super resolution algorithms can recover the high frequency loss to achieve high-fidelity audio, they require considerably more computational resources than available in low-power hearable devices. This paper proposes the first-ever real-time on-chip speech audio super resolution system for BCM. To accomplish this, we built and compared a series of lightweight audio super resolution deep learning models. Among all these models, \networkname is the most cost-efficient because the proposed novel Audio Temporal Shift Module (ATSM) reduces the network's dimensionality while maintaining sufficient temporal features from speech audios. Then we quantized and deployed the \networkname to low-end ARM micro-controller units for real-time embedded prototype. Evaluation results show that our system achieved real-time inference speed on Cortex-M7 and higher quality than the baseline audio super resolution method. Finally, we conducted a user study with ten experts and ten amateur listeners to evaluate our method's effectiveness to human ears. Both groups perceived a significantly higher speech quality with our method when compared to the solutions with the original BCM or air conduction microphone with cutting-edge noise reduction algorithms.

\end{abstract}



\maketitle

\section{Introduction}
\label{sec:intro}

The most commonly used microphones for voice communication are air conduction microphones, which pick up sound propagating through the air. Although providing high fidelity capture in quiet scenarios, they are vulnerable to environmental noises. To improve the speech quality of air conduction microphones in noise environments, researchers proposed multi-microphone beamforming with noise suppression techniques~\cite{capon1969high, pratt1972generalized, boll1979suppression} or deep learning based speech enhancement methods~\cite{park2016fully, macartney2018improved}. However, 
these solutions require a significant amount of additional hardware or computing resources. Further, all these methods fundamentally seek to reduce environmental noises but also inevitably corrupt speech. Moreover, these solutions are still vulnerable to boisterous environments such as construction sites or strong wind, where extraneous noises overpower speech signals.

Bone conduction microphones (BCMs) could achieve more robust results against ambient noises due to their physical design and fundamental conduction principles. However, the BCMs only have limited frequency response with high-frequency components above 2kHz significantly attenuated. Reconstructing the high-frequency details can effectively increase the speech audio's quality. A traditional method of reconstruction is to design a linear phase impulse response filter ~\cite{shimamura2005reconstruction}. However, acoustic paths are different among speakers because their bone structures are unique. Furthermore, it is impossible to ensure uniform BCM placement, which may result in different spectral properties~\cite{mcbride2011effect}. Therefore, a simple filter is insufficient to accommodate a variety of users. 

Audio super resolution~\cite{kuleshov2017audio}, also called bandwidth expansion, is the task of increasing audio sampling rate and restoring the high-frequency components of low-resolution audios. Convolutional Neural Networks have achieved state-of-the-art performance in audio super resolution~\cite{birnbaum2019temporal, kim2019bandwidth, hao2020time}. Additionally, similar neural network structures have also been proven effective in reconstructing distorted spectrograms~\cite{kegler2019deep} and enhancing recordings from low-end microphones~\cite{mysore2014can, su2020hifi}. Therefore, designing an audio super resolution model is feasible to reproduce high-fidelity speech from BCMs while maintaining their noise resistance property in multi-speaker settings. However, existing deep learning based audio super resolution methods are commonly computationally intensive, making them unfit for deployment on resource-constrained embedded systems. 

This paper proposes the first-ever real-time on-chip speech audio super resolution system for BCMs. In order to achieve this goal, we first designed and compared a series of lightweight deep learning models for speech audio super resolution. Among all the models, \networkname is the most cost-efficient. We proposed an audio temporal shift module (ATSM) and introduced this module to \networkname. Therefore, \networkname can reduce the network to one dimension but still learn enough features from the temporal information flow in speech audios. Thus, \networkname can reconstruct high-fidelity speech audios but require minimal computational resources. We further quantized and deployed \networkname and its variants on micro-controllers, including ARM Cortex-M4f and M7 processors, and conducted a full evaluation of our proposed method's performance in audio quality and inference latency. Results show that \networkname  outperforms the cutting-edge audio super resolution method~\cite{birnbaum2019temporal} with the perceptual evaluation of speech quality (PESQ)~\cite{recommendation2001perceptual} increased by 9\% and log-spectral distance (LSD)~\cite{gray1976distance} by 44\%. On the Cortex-M7 processor, our end-to-end latency, comprising model inference, feature extraction, and reconstruction, is 38 ms on average. This is less than half-frame length (64 ms), meaning our system can achieve real-time processing with 128 ms frames half-overlapped. To further assess our method's effectiveness in obtaining high quality speech, we recruited 20 participants, including 10 experts and 10 amateur listeners, for the perceptual audio quality evaluation. Results show that our method outperforms the original BCM solution and commodity noise reduction solution with the air conduction microphone. To the best of our knowledge, our method is the first chip deployable audio super resolution solution.
\textbf{To summarize, our contributions are as follows:}

\begin{itemize}

\item We proposed a lightweight audio super resolution deep learning model --- \networkname, which utilizes our proposed audio temporal shift module (ATSM) to form a novel one-dimensional UNet architecture. When compared with \networkname's variants without ATSM, \networkname is most cost-efficient for chip deployment. 

\item We implemented the first-ever real-time on-chip speech audio super resolution system for the bone conduction microphone by quantizing and deploying \networkname to popular micro-controllers in commodity hearable devices. We further evaluated its computational complexity on both ARM Cortex-M4f and M7 processors and demonstrated its real-time processing capability.



\item We evaluated our system's effectiveness in improving speech quality with a bone conduction microphone through perceptual audio quality user studies. Audio samples are publicly available~\footnote{\url{https://sites.google.com/view/audio-sr-for-bcm/home}}.

\end{itemize}

\section{Background and Related Work}
\label{sec:literature}
Voice communication using air conduction microphones in noisy environments has been a challenging problem. For conventional speech communication, researchers proposed speech enhancement methods, such as beamforming with a microphone array and blind source separation~\cite{hidri2012multichannel}, for background noise removal. These algorithms only remove part of the unwanted noises and introduce the risk of damaging the voice integrity. 
Beamforming is based on directionality. Therefore, it is prone to directional noise sources. In other words, when noise and voice sources are on-axis, beamforming will not effectively separate the noise. To reduce on-axis noise, noise suppression algorithms such as~\cite{boll1979suppression, ephraim1984speech, scalart1996speech} first estimate the noise with statistical models then remove the noise from the captured spectrum to recover the original speech. These methods could lead to speech integrity issues due to the noise model estimation accuracy. Moreover, under extreme conditions like strong wind noise, air conduction microphones will not pick up human voices due to saturation.

Bone conduction microphone, which collects human speech propagated via human bones, naturally suppresses environmental noises with its hardware placement and FSV conduction. However, the speech captured by the BCM has a limited frequency bandwidth which attenuates quickly above 2kHz~\cite{shin2012survey}. Our motivation is to enhance the BCM's speech sound quality by recovering high-frequency details while keeping its advantage against environmental noise. In this section, we describe existing speech enhancement algorithms for BCM and then give an overview on speech super resolution techniques.

\subsection{Bone Conduction Microphones}
\label{subsec:bcm}

Bone conduction microphones were commonly used as an accessorial enhancer to the air conduction microphones for capturing human speech. Researchers have proposed speech enhancement methods using BCMs~\cite{liu2004direct, lee2018bone, takada2018self, zhou2020real, yu2020time}. The BCMs can be used for accurate voice activity detection due to their noise suppression characteristics~\cite{zhou2020real}. The BCMs can also be incorporated to increase the voice activity detector accuracy and hence increase the accuracy for noise model estimation to achieve better denoising results~\cite{lee2018bone}. Further, BCMs can provide additional input for a multi-modal deep learning network~\cite{yu2020time}. However, these solutions require multiple microphones that are costly and limited in capability in extreme circumstances such as strong wind noise. Our work aims to enhance speech quality using a single bone conduction microphone. In other words, we plan to achieve clean human speech capture while maintaining the microphone's capability against environmental noises.

Similar speech processing techniques based on BCMs speech capture with audio super resolution can be found, including the following: speech enhancement approaches for bone conduction microphones through audio signal processing~\cite{shin2012survey}. \citet{shimamura2005reconstruction} proposed a reconstruction filter calculated from long-term spectra of human voices from both the air and bone conduction microphones.  \citet{shimamura2006improving} further utilized a multi-layer perceptron to model the reconstruction filter more accurately. \citet{rahman2011intelligibility} excluded the need for the air conduction microphone by introducing an analysis-synthesis method based on linear prediction. \citet{bouserhal2017ear} introduced adaptive filtering along with the non-linear bandwidth extension method for enhancing the speech sound quality.  However, these methods require complex feature engineering, thus difficult to adapt to different users and equipment setups.

Recently, researchers applied deep learning methods for speech enhancement with BCMs. These methods aim to increase the sound quality of BCMs to be comparable to air conduction microphones in ideal conditions. For example, \citet{shan2018novel} proposed an encoder-decoder network with the long short-term memory (LSTM) layer and local attention mechanism, which reconstructs an air conduction log-spectrogram from a bone conduction log-spectrogram. This method only reconstructs frequency components below 4kHz and is based on a specific speaker. \citet{liu2018bone} introduced Mel-scale features of speech audios from the bone conduction microphone with a deep denoise auto-encoder for speech enhancement. This work reconstructs high-frequency components up to 8kHz. It increases the perceptual evaluation of speech quality (PESQ) by 9.38 \% compared with the original bone conduction speech, but the auto-encoder is also trained with a single speaker's speech. \citet{hussain2019bone} proved that with only limited training data, the hierarchical extreme learning machine could outperform the denoise auto-encoder. \citet{zheng2019improving} adapted structural similarity (SSIM), a widely used metric in image quality assessment, as the loss function for a Bidirectional LSTM Neural Network. As a result, the model achieves higher PESQ when trained with SSIM loss compared with standard mean square error (MSE).

Although proven effective, the aforementioned deep learning methods were not designed to be deployed on real-time embedded systems due to exceeding computation resources and power limits. Moreover, these methods were not evaluated with cross-user validation, which limits their generalizability to adapt to individual users. Some works~\cite{shan2018novel, zheng2019improving} use the sampling rate of 8kHz, which is not sufficient for the Wideband Speech protocol with required sampling frequency at 16kHz. Therefore, our work approaches BCM voice capture as a real-time resource-constrained super resolution problem on embedded systems. Furthermore, to make our solution robust against individual users and various environments, we also introduced transfer learning to make our model generalizable.

\subsection{Audio Super Resolution Techniques}
\label{subsec:audioSR}
The audio super resolution, also known as bandwidth expansion, aims to increase the sampling rate and restore high-frequency components of the low-resolution audio. Inspired by the image inpainting methods, researchers have proposed several frequency domain based deep learning methods for audio super resolution. These methods can be trained using clean samples of BCMs as input and air conduction microphones as references. These samples are then converted into snapshots of spectrograms in the frequency domain as snippets of audio features. The learned model restores the missing high-frequency details from BCMs based on pattern recognition at inference time. Then the output snippets are reconstructed back into the real-time speech stream as output. Below we describe various audio super resolution methods, optimization strategies, and on-chip deployment methods. 

Audio super resolution methods either took the raw waveforms~\cite{kuleshov2017audio, birnbaum2019temporal, kim2019bandwidth, hao2020time} or spectral representations~\cite{li2015deep, eskimez2019adversarial, lim2018time} as the input. One-dimensional UNet~\cite{kuleshov2017audio}, asymmetrical network with skip connections, was the first attempt to use a deep convolutional neural network for end-to-end speech super resolution. To expand the perceptual field, TFiLM~\cite{birnbaum2019temporal} utilized bidirectional LSTM as the module to build up a variant 1D-UNet for speech audio super resolution. In another variation of 1D-UNet~\cite{kim2019bandwidth}, conventional convolutional layers were replaced by multi-scale convolutional layers to capture information at multiple scales. Mfnet~\cite{hao2020time} also tried to facilitate multi-scale information exchange through multi-scale fusion block. Other deep learning methods utilized the spectrogram as input. For example, \citet{li2015deep} proposed a three-layer fully connected network for speech audio super resolution. UNet~\cite{eskimez2019adversarial} was also proven to be effective in performing speech audio super resolution using the power spectrogram. To take advantage of representations in both time and frequency domain, TFnet~\cite{lim2018time} incorporated two network branches that operate on both the waveform and the spectrogram, respectively. Although proven effective on speech audio super resolution, the deep learning models mentioned above have too many parameters causing them to exceed the computation and power budgets of micro-controllers by a factor of 100 times.

The optimization strategy for audio super resolution includes the loss function and training optimizations. One of the most commonly used loss functions is the mean square error~\cite{kuleshov2017audio, li2015deep}. Although simple to compute, the MSE does not represent the human perceptual speech quality. Therefore, perceptually motivated loss~\cite{feng2019learning} that calculates L1 distance on log mel-spectrogram was proposed. Further, the log spectral distance (LSD)~\cite{gray1976distance, kuleshov2017audio}, which measures the distance between log-power spectrum of reference and reconstructed signals, was also adopted as one option for the loss function. For training, WaveNet~\cite{wang2018speech, gupta2019speech}, an auto-regressive model, optimized the joint probability of the targeted high-resolution audio. Adversarial learning is another popular training technique. In this technique, a discriminator that works either in the time domain \cite{kim2019bandwidth, hao2020time, li2020real} or frequency domain \cite{eskimez2019adversarial, li2019speech, kumar2020nu} guides the generator to predict more realistic high-resolution audio from low-resolution inputs. 

Recently, hearables devices, such as TWS earbuds, have become increasingly popular, with 233 million shipments in 2020. While deep learning based audio super resolution methods have been proven effective, deploying such solutions to a resource-limited embedded system has not been fully investigated. Similar to our proposal, several super resolution deep learning methods~\cite{li2020real, liu2021splitsr, lee2019mobisr} have proven the feasibility of applying the super resolution method on a smartphone. Other state-of-the-art speech super resolution models require considerable computation resources and cause significant latency, which is not suitable for edge device deployment. This paper proposed a lightweight deep learning model --- \networkname, which can run on power- and space-limited ARM Cortex-M platforms. We expect future hearables embedded with a single BCM will achieve good performance without the need for additional computation resources with the proposed method.

\begin{figure}[ht]
  \includegraphics[width=\linewidth]{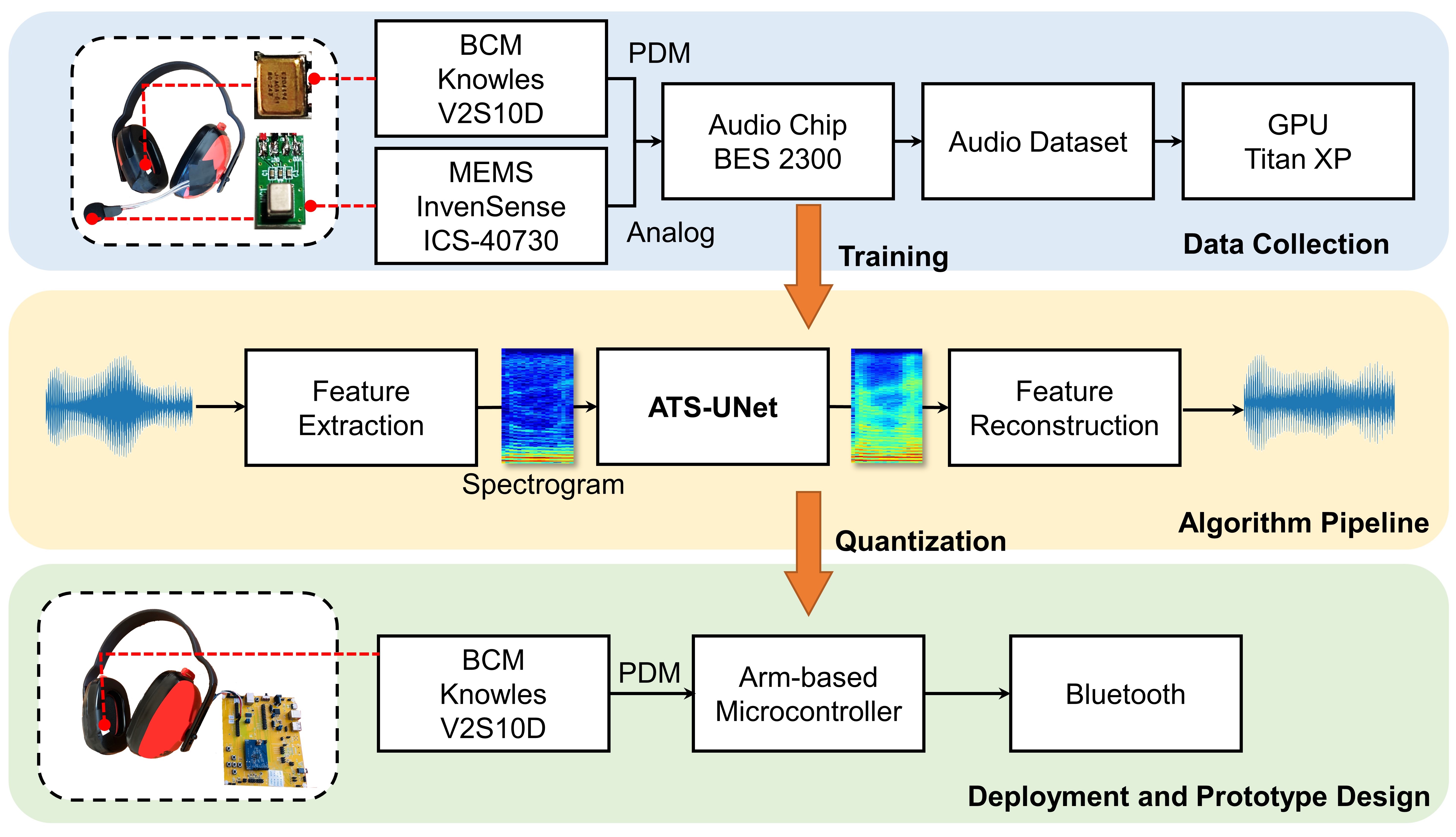}
  \caption{This paper's overview.}
  \Description{}
  \label{fig:system}
\end{figure}

\section{Overview}
\label{sec:SD}

This paper focuses on the uplink portion of the communication system, namely the capture side of the speech communication protocol. In particular, the capture and recovery of the BCM input as an alternative solution to the conventional air conduction microphone. Shown in Figure~\ref{fig:system}, our proof-of-concept prototype is composed of commercially available electronic parts: a pulse density modulation (PDM) bone conduction microphone (Knowles V2S100D), an analog MEMS air conduction microphone (InvenSense ICS-40730), and a micro-controller development board (Bestechnic~\footnote{\url{http://www.bestechnic.com/Home/Index/index/lan_type/2}} (BES) BES2300YP). The BES2300YP system on chip (SoC) simultaneously collects audio signals from the Knowles V2S100D and InvenSense ICS-40730, forming a dataset for the audio super resolution. Then, we trained \networkname using this dataset on an Nvidia Titan XP GPU (12GB RAM). 
The floating-point model was further quantized to 16-bit data format and transformed from Tensorflow~\cite{abadi2016tensorflow} to CMSIS-NN~\cite{lai2018cmsis} implementation. With the model quantization and optimization,  \networkname can then run efficiently on micro-controllers. 

In this work, we tested two popular micro-controllers. 1) The BES2300YP with dual ARM-Cortex M4F processors operating at a frequency up to 300MHz with 992KB SRAM and 4MB Flash storage. The BES SoC was adopted by many popular TWS earbuds, such as JBL FREE II, Samsung Galaxy Buds Live, and Huawei FreeBuds 2 Pro, for its compact form factor and power efficiency. We only use one single processor in this work since the other processor runs the Bluetooth stack and digital signal processing (DSP) related algorithms. Furthermore, the two processors share the SRAM with the storage requirements from the Bluetooth and the operating system taking more than 400KB. To prevent the memory overflow, we limited the SRAM for the machine learning model to be below 500KB. 2)  The NXP RT1060 SoC with a single Arm-Cortex M7 processor operating at a frequency up to 600MHz with 1MB on-chip SRAM. In this case, only 512KB general-purpose SRAM can be used to host the machine learning model. Both micro-controllers support audio applications.

\section{Deep Learning Models for Bone Conduction Speech Audio Super Resolution}
\label{sec:ats}
In this section, we first describe our general UNet design for bone conduction speech audio super resolution. We then describe how our models, including our proposed 2D-UNet, Hybrid-UNet, Mixed-UNet, 1D-UNet, and \networkname, were derived from this UNet design. Most importantly, we present the key module called the Audio Temporal Shift Module (ATSM). Finally, we describe the pre-processing and post-processing methods for our deep learning models. 

\begin{figure}[ht]
  \includegraphics[width=0.8\linewidth]{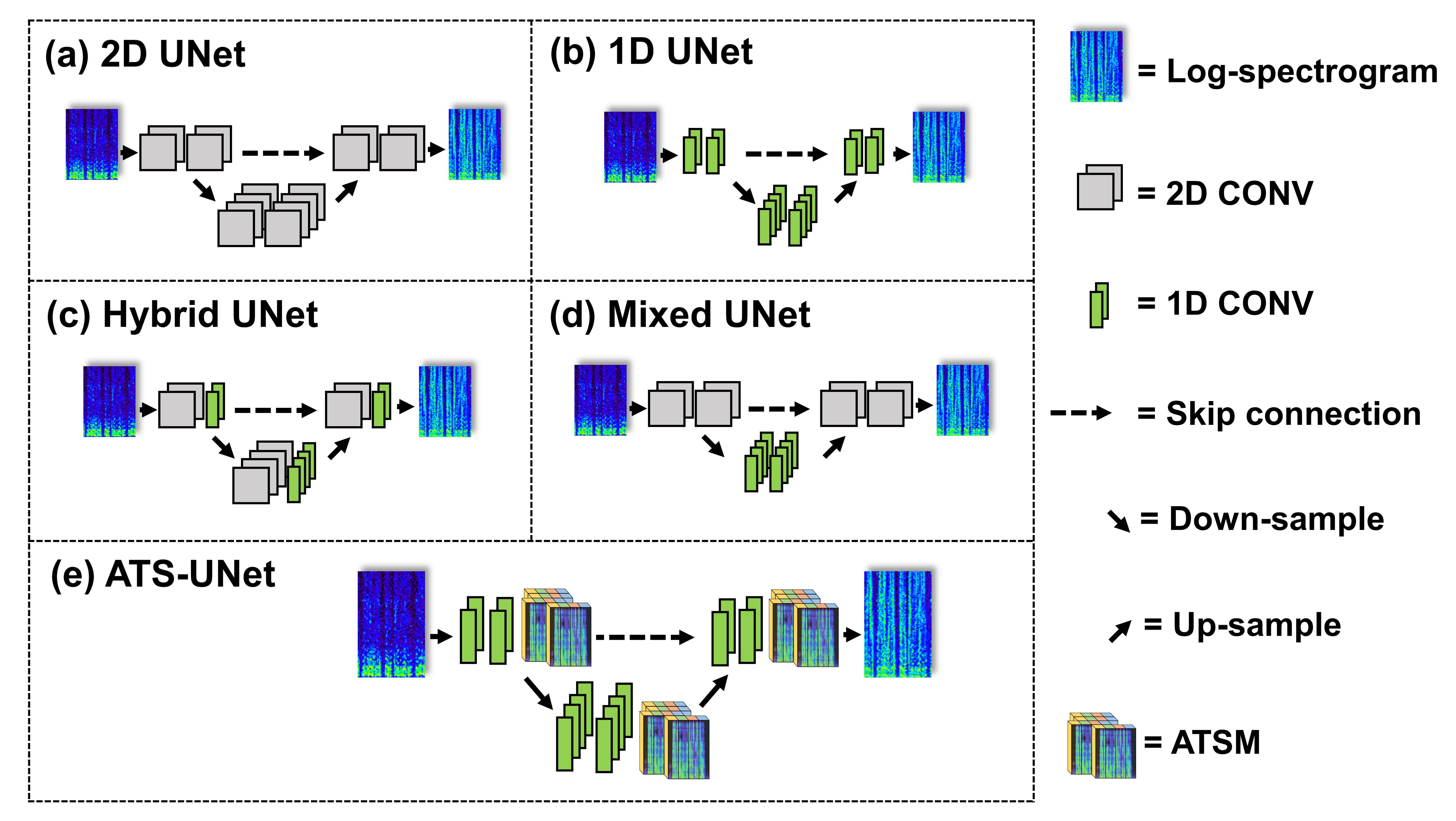}
  \caption{Our proposed series of novel network architectures including a variant 2D-UNet, 1D-UNet, Hybrid-UNet, Mixed-UNet, and \networkname}
  \label{fig:unet1}
\end{figure}

\subsection{UNet Variances for Bone Conduction Speech Audio Super Resolution}
\label{UNetbaseline}

\begin{figure}[ht]
  \includegraphics[width=\linewidth]{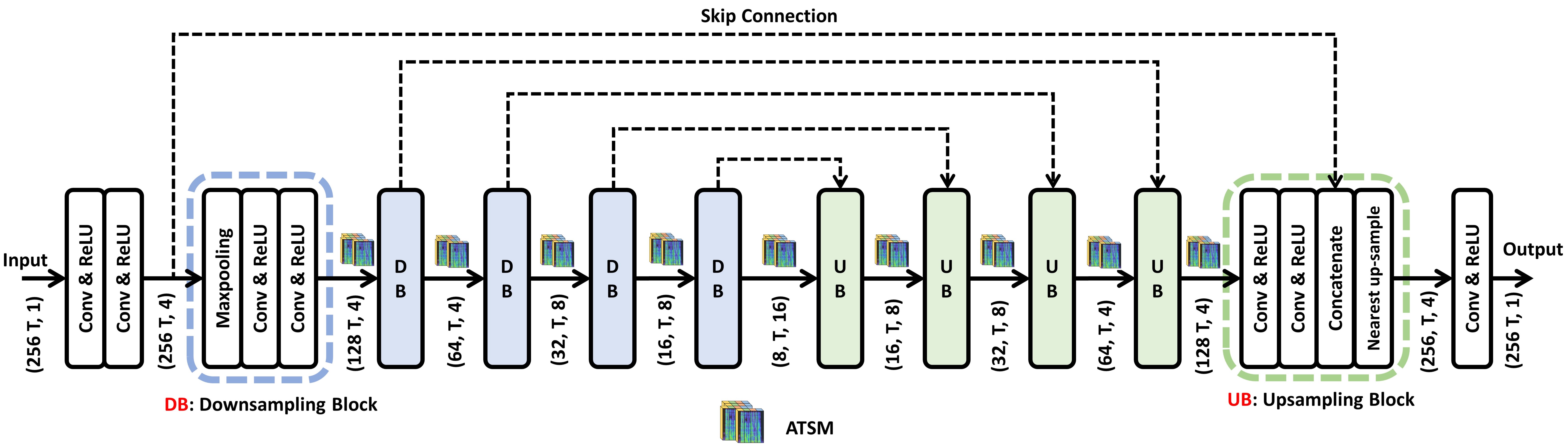}
  \caption{The detailed architecture of our general UNet design. $(F, T, C)$ stands for $F$ frequency bins, $T$ temporal frames and $C$ channels. When all convolutional layers are 1D and ATSMs are inserted after each DB/UB, it is our \networkname. }
  \Description{}
  \label{fig:unet2}
\end{figure}

The original UNet has a fully convolutional and symmetrical network structure with skip connections to facilitate information flow. Additionally, it can extract temporal and frequency information in the time-frequency domain and reconstruct high-resolution spectrograms. Compared with conventional convolutional and recurrent neural networks, UNet is more efficient, as feature maps are down-sampled, contributing to fewer floating-point operations. However, UNet's large model size still introduces unfavorable computation for on-chip deployment. Therefore, we reduce the number of channels and network depth. The general UNet architecture (Figure~\ref{fig:unet2}) contains 5 down-sampling blocks (DB) and 5 up-sampling blocks (UB). Each DB has a max-pooling layer followed by two convolutional layers. The size of the max-pooling layer is 2$\times$1. Therefore after each DB, the length of the frequency axis of the feature map is halved, while the time dimension remains the same throughout the network. In UB, feature maps are up-sampled, concatenated with skip features then fed into two convolutional layers. ReLU activation function is adopted after each convolutional layer except for the last layer. \textbf{2D-UNet V1} (Figure\ref{fig:unet1} a) has the same structure and channel numbers as shown in Figure.~\ref{fig:unet2}, but it has no ATSM and every convolutional layer is 2D. We also present \textbf{2D-UNet V2} which has 4 times the filters of 2D-UNet v1 for comparison.

Although 2D-UNet V1 is significantly smaller than the original UNet for image segmentation, 2D convolutional layers still introduce unfavorable computation for on-chip deployment. Since low-latency audio super resolution requires a small frame size, the shape of the input spectrogram is narrow (the frequency axis is much longer than the time axis). Therefore, only a few 2D convolutional layers are enough to extract information from the full temporal range. Thus using 2D convolutional layers throughout the network is unnecessary. 

Based on the above observation, We present another architecture called \textbf{Hybrid-UNet} (Figure\ref{fig:unet1} c) and \textbf{Mixed-UNet} (Figure\ref{fig:unet1} d), which replace a portion of 2D convolutional layers with 1D convolutional layers. 2D convolutional layers enable temporal information flow, while 1D convolutional layers only compute along frequency dimension to enlarge perceptual range.  Hybrid-UNet adopts 2D and 1D convolutional layers in each DB/UB alternately, which maintains temporal information flow in the whole network. Mixed-UNet replaces 2D convolutional layers to 1D in the middle of the network so that temporal information exchange only exits in shallow layers.  

Although Hybrid-UNet and Mixed-UNet are more efficient than traditional 2D-UNet, 2D convolutional layers still introduce unfavorable computation for real-time inference on low-end embedded systems. Thus, we replace 2D convolutional layers with 1D convolutional layers completely to get a new architecture called \textbf{1D-UNet} (Figure\ref{fig:unet1} b), 1D version of 2D-UNet V1. However, 1D-UNet lacks temporal modeling, so we inserted ATSM after each DB/UB to enable efficient and effective information exchange along the temporal axis. We called 1D-UNet with ATSM \textbf{ATS-UNet} (Figure\ref{fig:unet1} e). For all the models, the kernel sizes of 1D and 2D convolutional layers are 3$\times$1 and 3$\times$3, respectively.

\subsection{Audio Temporal Shift Module (ATSM)}

\begin{figure}[ht]
  \includegraphics[width=0.7\linewidth]{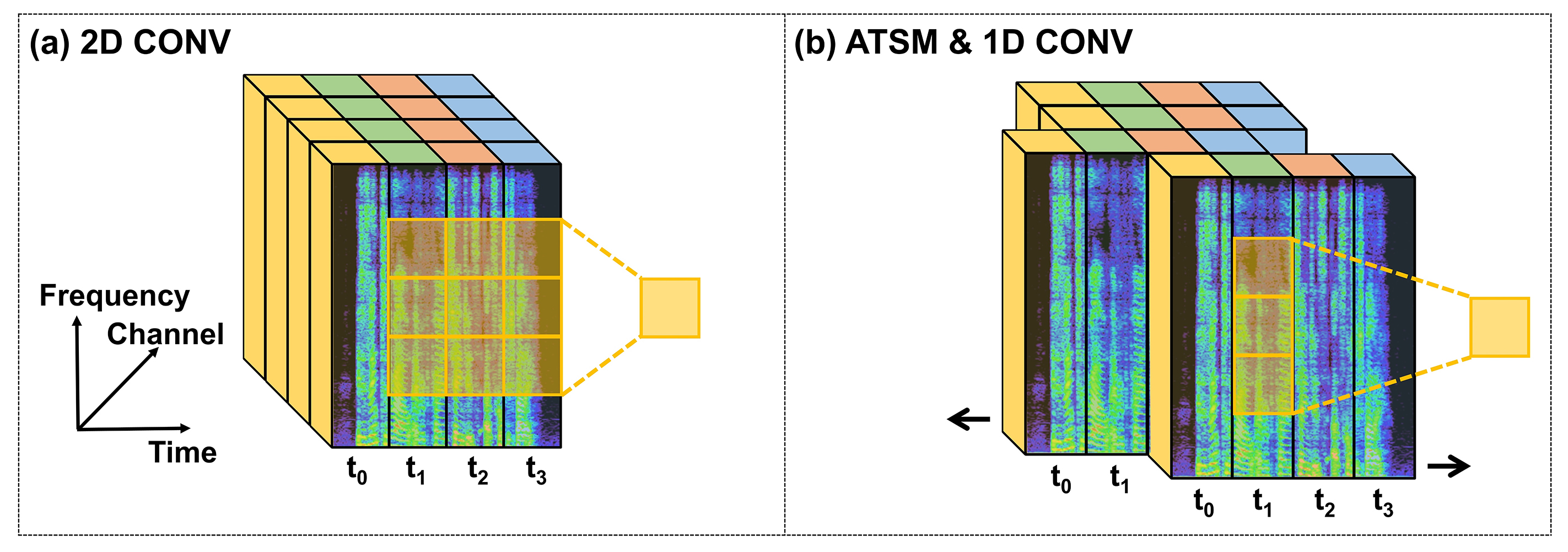}
  \caption{To enable information flow through $t_{1}$, $t_{2}$, $t_{3}$, we can use (a) a 2D convolutional layer or (b) the proposed ATSM with 1D convolutional layer. The latter one is more lightweight.}
  \Description{}
  \label{fig:tsm}
\end{figure}

Conventional deep learning models for audio processing require massive 2D convolutional operations to extract meaningful features from spectrograms~\cite{park2016fully} as Figure~\ref{fig:tsm}(a) shows. However, they utilize a large number of computational resources. Therefore, the aforementioned deep learning models are unlikely to be adopted for on-chip audio super resolution. Instead, we introduced a novel module to accelerate convolutional operations in the time-frequency domain called the Audio Temporal Shift Module (ATSM), as Figure~\ref{fig:tsm}(b) shows. ATSM was inspired by the Temporal Shift Module (TSM)~\cite{lin2019tsm}, an effective mechanism for video understanding. It replaces 3D convolutional operations with 2D ones while preserving high-dimensional modeling. This is achieved by shifting the feature maps among video frames to enable temporal information flow. Similarly, ATSM utilizes 1D convolution operations to replace 2D convolution operations for audio processing. In order to utilize information from a longer temporal range for 1D convolutional kernels, ATSM shifts feature maps along the temporal axis of spectrograms. In contrast to the TSM, whose input is four-dimensional feature maps extracted from video frames, the input ATSM is extracted from the 2D log-spectrogram, only has three dimensions: channel, time, and frequency. 

More specifically, as illustrated in Figure~\ref{fig:tsm}(b), feature maps are divided into two chunks along the channel dimension: 1) dynamic and 2) static. The dynamic feature maps are split evenly into two parts, with one part shifted forward  (delaying time by one frame) and the other backward (advancing time by one frame). The static feature maps remain unchanged. It is worth noting that ATSM requires no additional computational resource to facilitate information exchange along the temporal dimension in spectrogram computation.

\subsection{Audio Pre-processing and Resynthesis}
\label{subsec:FER}

Voice communication's ideal overall latency is below 50 ms that humans are unable to notice. As latency increases, humans start to notice lip-sync issues, but communication latency under 150ms is still considered acceptable. However, a latency that exceeds 400ms~\cite{abbas1996itu} is unacceptable for real-time communication. Therefore, a feasible audio super resolution system should not add too much latency to the communication process. As a result, our system requires fast computing with appropriate frame size and short-time Fourier transform (STFT) parameter.

\begin{figure}[ht]
  \includegraphics[width=0.9\linewidth]{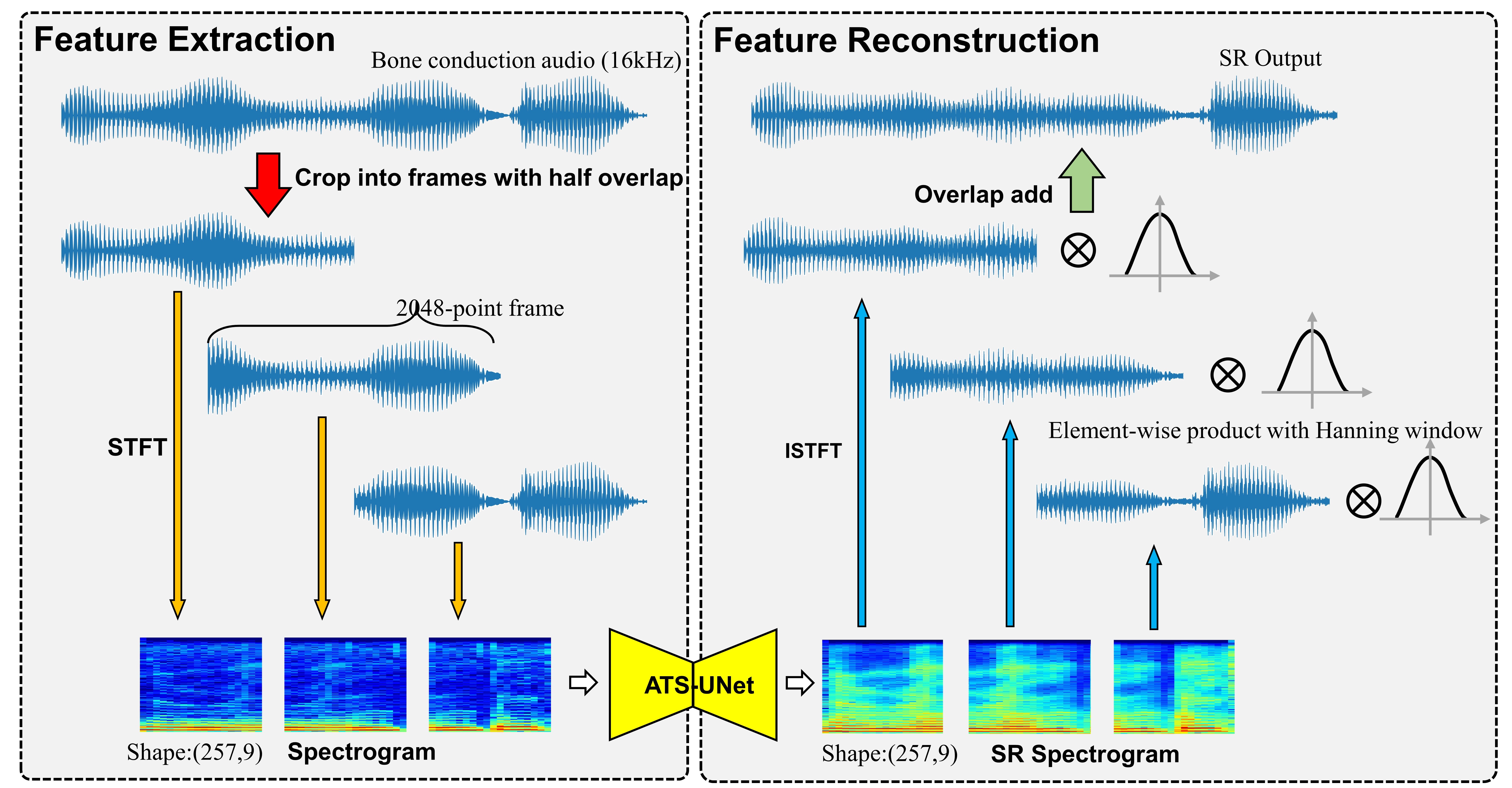}
  \caption{Feature Extraction and Reconstruction}
  \Description{}
  \label{fig:feature}
\end{figure}

The pre-processing includes the feature extraction from the raw audio signal as shown in the left figure of Figure~\ref{fig:feature}. \networkname processes a single audio frame at a time and outputs audio frames in sequence to resynthesize the audio stream. A large frame provides more information for \networkname but introduces longer latency, since the system has to wait for the time of the entire frame. Therefore, we use a 2048-point (128ms) frame with half overlap to achieve acceptable latency while maintaining adequate information. These frames are transformed into spectrograms by STFT and fed into \networkname. 

The STFT parameter is another major factor in computational intensity. High frequency and time resolution spectrograms can be achieved with a larger fast Fourier transform (FFT) size and overlap between FFT windows, resulting in considerable computation load. Considering the memory and resources on the micro-controller, we adopted a window size of 512 points for the STFT. Further, we also utilized a half overlap strategy to the raw audio data. A detailed trade-off of the STFT parameter is explained in section~\ref{sec:evaluation}.

The audio resynthesis, also known as post-processing, converts the reconstructed spectrograms back to the time domain using the inverse short-time Fourier transform (ISTFT). The overlapped frame is then multiplied by the Hanning window (2048-point) and added to the previous frame. We adopted this resynthesis method because the data points in the center of the window are better reconstructed due to richer temporal information. Therefore, the Hanning window function gives the data samples in the center of the window higher weights but weakened the importance of the data samples by the side. Further, we adopted half-overlapped Hanning window functions so that their summation is a constant value. Thus it won't distort the signal and can smooth the transition between adjacent frames.

\section{Model Training and Deployment}
\label{sec:evaluation}
In this section, we provide experimental details, including data collection procedure, training scheme, and quantization procedure. 

\subsection{Speech Audio Data Collection}
\label{subsec:SADC}
\begin{figure}[ht]
  \includegraphics[width=0.65\linewidth]{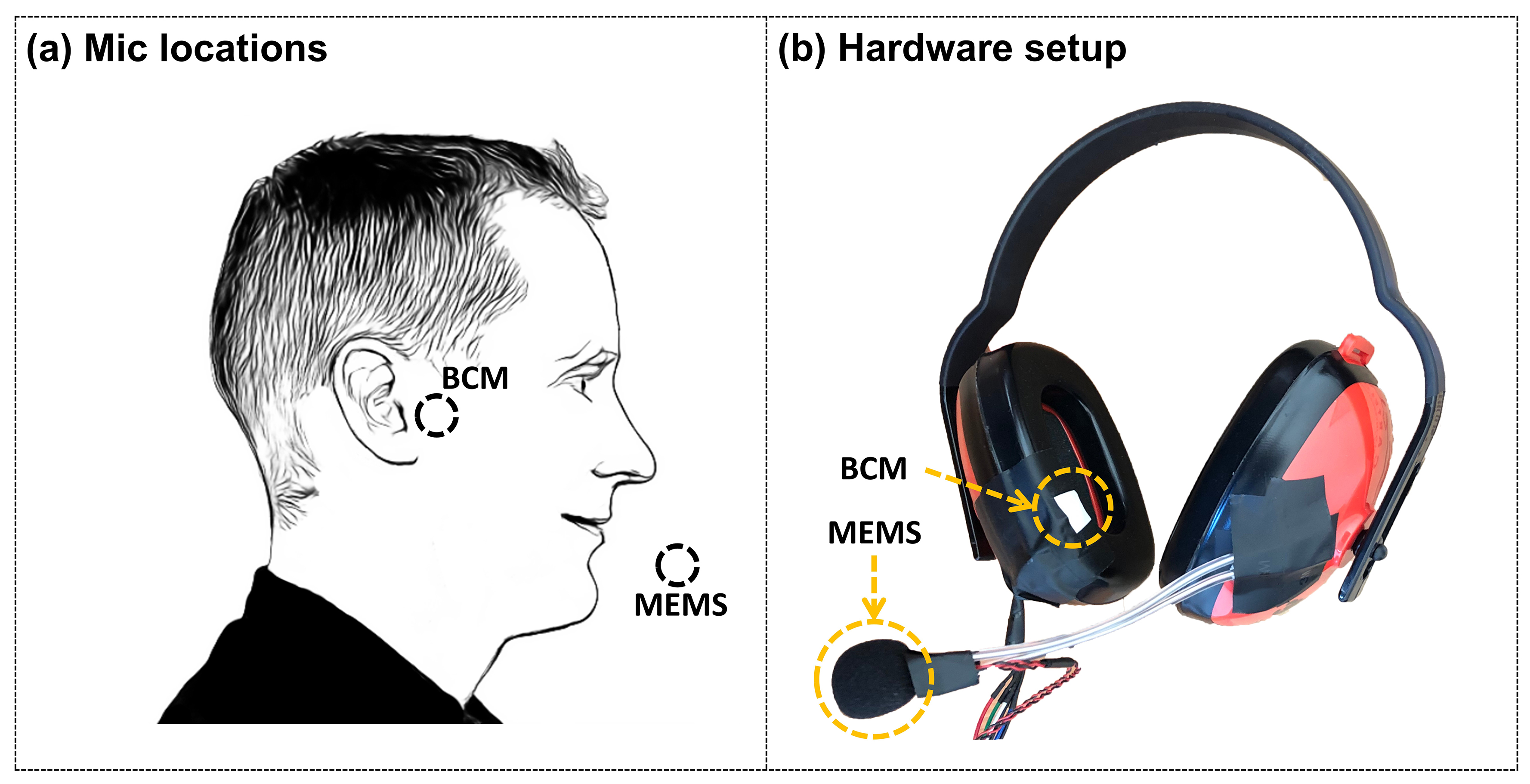}
  \caption{(a) The position of the air (MEMS) and the bone (BCM) conduction microphone.  (b) The headphone prototype for data collection.}
  \Description{}
  \label{fig:mic}
\end{figure}

We conducted a user experiment to collected an audio dataset using the hardware shown in Figure~\ref{fig:mic}. The MEMS air conduction microphone was placed near the mouth to collect high-quality ground-truth speech audios. The BCM was secured with an earmuff. So when subjects put on the earmuff, BCM would be pressed in front of the ear. To prevent reverberation, we put an acoustic panel in front of the speaker. As Figure~\ref{fig:system} indicates, we utilized a BES2300YP micro-controller to simultaneously collected speech audios from both the air and bone conduction microphones. The sampling rate and bit depth were set to 44.1 kHz and 16 bits. We recorded the speech audios in a recording studio that is quiet for high speech quality. 

We recruited 20 participants (10 males and 10 females). After wore the headphone, each subject was informed to read six paragraphs of an article, yielding approximately 12 minutes of speech per subject. We removed the silence clips at the beginning and the end of all audio files and normalized the volume across participants. We then down-sampled each speech audio to 16 kHz, which is sufficient for communication~\footnote{\url{https://en.wikipedia.org/wiki/Sampling_(signal_processing)}}. The processed dataset includes 200 minutes of speech audios in total. Each participant received a 10 USD gift card after the experiment. 


\subsection{Implementation and Training Details}
\label{subsec:ITD}
Bone conduction audios were down-sampled to 16 kHz, cropped to 2048-point frames (128ms), and transformed to time-frequency representations by short-time Fourier transform (STFT) \cite{allen1977unified} with 512-point Hanning window and half overlap (a stride of 256-point). We adopted the implementation of librosa~\footnote{\url{https://librosa.org/doc/latest/index.html}} for STFT and ISTFT. The symmetrical component is removed, resulting in 257 Fourier coefficients. Thus, the size of the input spectrogram is 257$\times$9. Then, we converted the power of each coefficient to the log scale and standardize them to normal distribution. We skipped the $0^{th}$ coefficient (DC component) but fed the rest 256 coefficients into the network. Lastly, we obtained the enhanced log-spectrogram from the super resolution model and concatenated it with $0^{th}$ coefficient, so the output's shape is also 257$\times$9. The post-processing audio resynthesis includes denormalization, conversion to linear-scale, and inverse STFT. The model only predicts magnitude, so we kept the original phase information from bone conductive audio to resynthesize the enhanced speech audio.

All super resolution models were implemented in Tensorflow~\cite{abadi2016tensorflow}. We adopted 10-fold cross user validation with the training dataset consisting of speech audios from 18 speakers and the rest audios as the test dataset. We randomly initialized the model and trained it for 100 epochs, using the Adam optimizer~\cite{kingma2014adam}, with a learning rate of 0.0001 and batch size of 64. 

\subsection{Loss Function}
\label{subsec:loss_fun}
The loss function is given by Equation~\ref{eq:loss}, 
that consists of two parts: least absolute deviation (L1) loss and perceptually motivated loss~\cite{feng2019learning}. L1 loss measures the absolute difference between the log-spectrograms of the output speech audio and the ground-truth speech audio --- $\log(s(y))$. Perceptually motivated loss is the L1 distance calculated on log-melspectrogram $\log(ms(y))$ considering that the mel-scale is more aligned with human hearing~\cite{Volkmann_1937}. In Equation~\ref{eq:loss}, $s(y)$ and $s(\hat{y})$ stand for spectrograms of the output and ground-truth audios. $ms(y)$ and $ms(\hat{y})$ represent melspectrograms of the output and ground-truth audios respectively. 

\begin{equation}\begin{aligned}\label{eq:loss}
 Loss=|\log(s(y))-\log(s(\hat{y}))|_{1}+
 |\log(ms(y))-\log(ms(\hat{y}))|_{1}
\end{aligned}\end{equation}

\subsection{Model Quantization}

We re-compiled each model using CMSIS-NN \cite{lai2018cmsis} framework for efficient inference on Arm Cortex-M processors. First, we transformed the model from floating-point to fixed-point format. Both weights and activations were quantized to 16-bit integers, given by Equation~\ref{eq:quantize}. In practice, quantization is symmetrical around zero with power-of-two scaling, so it can be implemented by bitwise shifts in CMSIS-NN kernels.

\begin{equation} \label{eq:quantize}
 x_{q}=\lfloor{x\times 2^{15-\log_2{max|x|}}}\rfloor
\end{equation}
where $x$ represents weights/activations of a convolutional layer. $x_q$ is quantized weights/activations.

\subsection{Noise Transfer Learning}
\label{subsec:n}
The primary motivation behind the use of BCM is to enhance communication quality in noisy environments. Therefore, our system should improve bone conductive recordings in a quiet environment and in boisterous environments. To this end, we collected voices from BCM in different noisy locations. However, we observed more unwanted noises in the reconstructed speech compared with the quiet laboratory setup. The reason is that BCM can still pick up some external noises and further being enhanced by the model. 

We adopted transfer learning to fine-tune the model for noise reduction. By collecting pure noises using BCM and adding them to bone conductive audio in the training dataset, the model can learn to identify the unwanted noises and only recover speech signals. In detail, we collected bone conductive noises in three locations, including the subway station, the bus stop, and the dining hall. We instructed a participant to wear the prototype without speaking and recorded bone conductive noises for 20 minutes in each location. We then added the noises to the bone conduction speech recorded in the quiet studio. For each audio clip, the signal-to-noise ratio (SNR) between the bone conductive speech and the additive noise was randomly sampled from Gaussian distribution with a mean of 18 and a standard deviation of 3.5. 
Before being deployed in real-world environments, \networkname was fine tunned on the noisy data for another 100 epochs with the same parameters in section~\ref{subsec:ITD}.

\section{Quantitative Speech Quality Evaluation}
\label{sec:AR}

In this section, we present the quantitative speech quality evaluation regarding the air conduction microphone as the ground truth. We describe the evaluation metrics, baselines, and results. We then explain the reasons behind the hyper-parameter selection and benchmark the performance of \networkname, UNet variants, and baselines for speech enhancement.

\subsection{Evaluation Metrics}
\label{subsec:metric}
We considered the effectiveness, model size, latency, and power consumption to evaluate each model's performance comprehensively. Specifically, the effectiveness includes two metrics: the log spectral distance (LSD)~\cite{gray1976distance}, and the perceptual evaluation of speech quality (PESQ)~\cite{recommendation2001perceptual}. LSD, given by Equation~\ref{eq:lsd}~\cite{kuleshov2017audio}, measures the distance between the log-power spectrum of reference and reconstructed signals. Therefore, a lower value indicates a better performance. PESQ was provided by Recommendation ITU-T P862~\cite{recommendation2001perceptual} for objective assessment of speech quality. It models mean opinion score (MOS) that ranges from 1 (bad) to 5 (excellent). 

\begin{equation} \label{eq:lsd}
LSD(x, \hat{x}) = \frac{1}{T}\sum_{t=1}^{T}\sqrt{\frac{1}{K}\sum_{k=1}
^{K}(X(t, k)-\hat{X}(t, k))^2}
\end{equation}
where $t$ and $k$ are frame and frequency index respectively. $X$ and $\hat{X}$ denote log-power spectrum of $x$ and $\hat{x}$, which are defined as $X=log|S(x)|^{2}$. $S$ stands for STFT with 2048-point frames.

\subsection{Baselines}

\citet{birnbaum2019temporal} inserted temporal feature-wise linear modulation (TFiLM) layers into a time-domain 1D-UNet to expand the receptive field. It improved the performance of audio super resolution compared with the original 1D-UNet~\cite{kuleshov2017audio}, achieving cutting edge audio super resolution performance. Therefore, we adopted TFiLM as the baseline in this paper. We used the open-sourced code of TFiLM implementation ~\footnote{https://github.com/kuleshov/audio-super-res}. 
 
\begin{table}[ht]
\centering
\begin{tabular}{|c c | c c|} 
\hline
\multicolumn{2}{|l|}{}  & \multicolumn{2}{c|}{Average LSD(dB) / PESQ}      \\ 
 \hline
 Model & Params & 1024-point STFT & 512-point STFT \\ 
 \hline
2D-UNet v1 & 11.8k & 2.028 / 2.713 & 2.013 / 2.790\\
2D-UNet v2 & 187.3k & 1.949 / 2.937 & 1.961 / 2.983\\
TFiLM (baseline)~\cite{birnbaum2019temporal} & 68221.2k & \multicolumn{2}{c|}{3.646 / 2.523 (time domain)}\\ 
 \hline
\end{tabular}
\caption{Performance comparison for different STFT parameter}
\label{table:stft}
\end{table}

\subsection{Effect of Input Hyper-parameter}
To evaluate the trade-off between frequency resolution and model performance, we first compared two sets of STFT parameters: 1024-point FFT, 256 strides, Blackman window, and 512-point FFT, 256 strides, Hanning window. Experiments were performed on two models. The first model is a lightweight 2D-UNet v1. In the second model, we expanded 2D-UNet v1 by increasing the number of filters in each layer by four times to explore optimum audio super resolution results without considering computation. As shown in Table~\ref{table:stft}, both 2D-UNet v1 and v2 outperform the baseline method --- TFiLM~\cite{birnbaum2019temporal} with significantly fewer parameters. This proves the effectiveness of 2D-UNet in speech audio super resolution. We noticed that both LSD and PESQ slightly improved with a larger window size for STFT. However, the computational intensity of the 1024-point STFT is nearly doubled compared with the 512-point STFT. Therefore, we adopted a 512-point window size for the STFT in the following evaluation procedures considering the latency and model size. 

\begin{table}[ht]
\centering
\begin{tabular}{|c c c | c c| c c|} 
\hline
\multicolumn{3}{|l|}{} & \multicolumn{2}{c|}{Latency(ms)}  & \multicolumn{2}{c|}{Average LSD(dB) / PESQ}      \\ 
 \hline
 Our Models & Params & FLOPs & Cortex-M4f & Cortex-M7 & L1 & L1+Perceptual Loss \\ 
 \hline
2D-UNet v2 & 187.3k & 133.9M & / & / & 1.961 / 2.983 & 1.954 / 3.030\\ 
2D-UNet v1 & 11.8k & 8.6M & 187 & 44 & 2.013 / 2.790 & 2.004 / 2.780\\
Hybrid-UNet & 8.4k & 7.0M & 163 & 38 & 2.024 / 2.689 & 2.015 /2.733\\
Mixed-UNet & 6.3k & 6.9M & 166 & 39 & 2.026 / 2.692 & 2.019 / 2.743\\
1D-UNet& 4.5k & 4.8M & 129 & 31 & 2.063 / 2.664 & 2.052 / 2.717\\
\textbf{\networkname} & 4.5k & 4.8M & 131 & 32 & 2.032 / 2.710 & 2.032 / 2.749\\
\hdashline
TFiLM (baseline)~\cite{birnbaum2019temporal} & 68221.2k & 116420M & / & / & \multicolumn{2}{c|}{3.646 / 2.523} \\
 \hline
\end{tabular}
\caption{Performance comparison for different UNet architectures. Latency is the model inference time on a single 2048-point frame by Arm Cortex-M4f/M7 processor.}
\label{table:result}
\end{table}

\subsection{Model Performance Result and Comparison}

To align the loss function with human hearing sensitivity for different frequency ranges, we incorporated perceptually motivated loss~\cite{feng2019learning}. Compared with L1 loss, it increases accuracy for every tested architecture (Table~\ref{table:result}).

\begin{figure}[ht]
  \includegraphics[width=\linewidth]{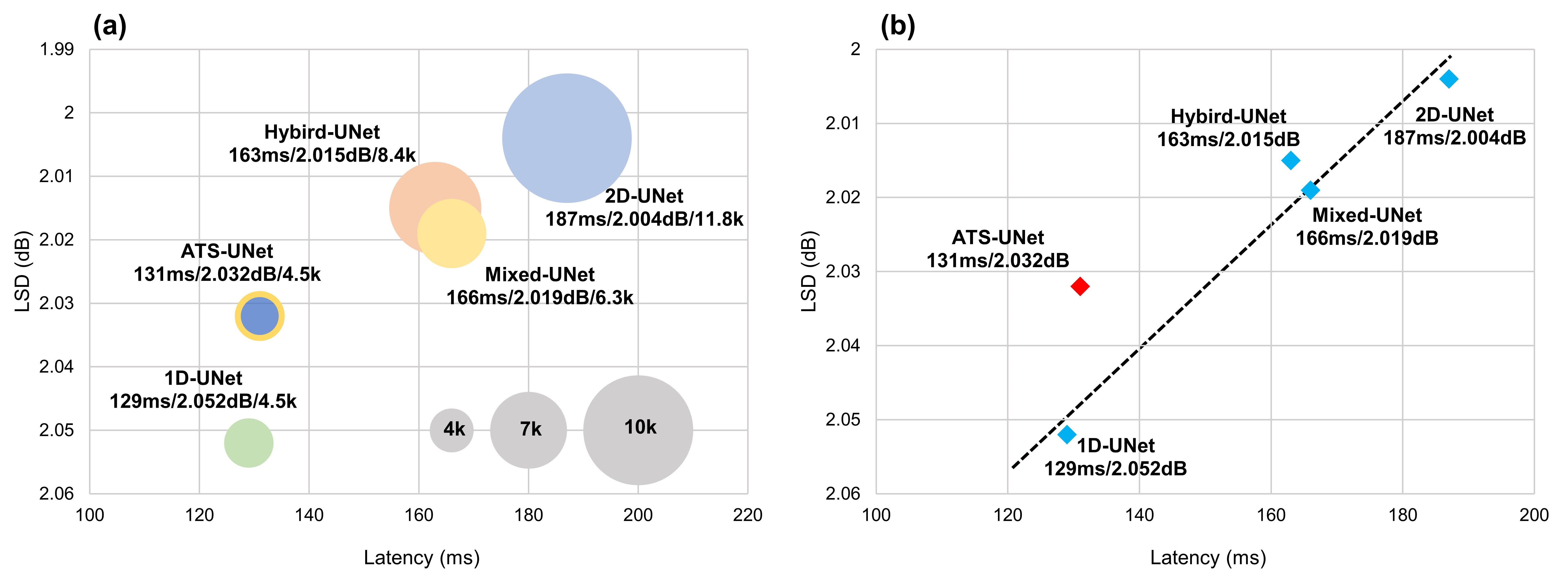}
  \caption{On-chip latency evaluation.}
  \Description{}
  \label{fig:bubble}
\end{figure}

Although UNet v1 only has about 10 thousand parameters, it still requires a long inference time on computation restricted platform, so we proposed Hybrid-UNet, Mixed-UNet, and \networkname, described in section~\ref{sec:ats}. Benchmark latencies and accuracies are provided in Table~\ref{table:result} and Fig.\ref{fig:bubble}. \networkname and 1D-UNet are the fastest networks, taking 131/32 ms and 129/31 ms respectively to inference a 2048-point frame. Due to the lack of temporal modeling, the accuracy of 1D-UNet is significantly lower than \networkname. ATSM effectively promotes temporal modeling while only adding negligible latency. Although Hybrid-UNet has 2k more parameters than Mixed-UNet, the two settings achieve nearly the same latency and accuracy because their floating-point operations (FLOPs) are very close. 2D-UNet v1 is the slowest with expensive computation and slightly higher accuracy. Note that 2D-UNet v2 is too large to be run on our embedded system, so we leave gaps in the table.

\begin{figure}[ht]
  \includegraphics[width=\linewidth]{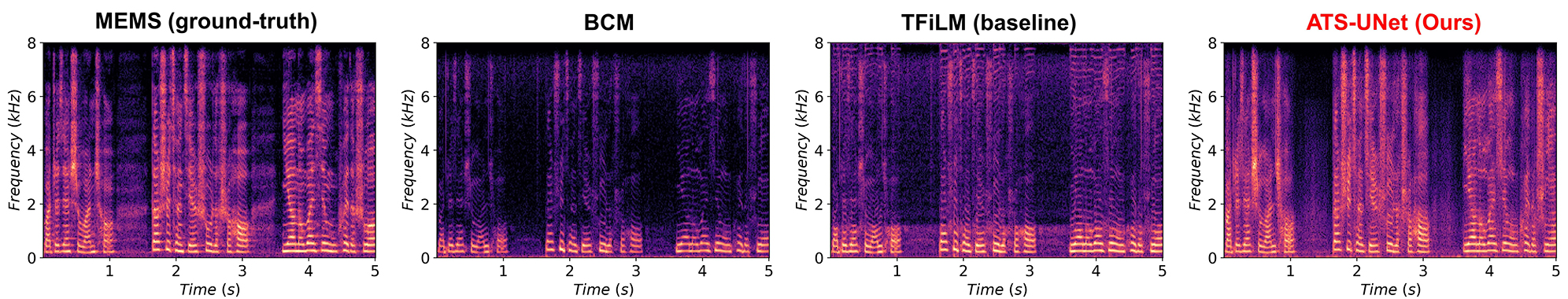}
  \caption{Audio super resolution results visualized by spectorgrams.}
  \Description{}
  \label{fig:spectrum}
\end{figure}

As shown in Figure~\ref{fig:bubble}, \networkname is the most cost-efficient model as it is on the upper left of the plot. Besides, spectrum examples in Figure~\ref{fig:spectrum} demonstrate \networkname outperforms TFiLM since it recovers a more accurate high-frequency structure. Considering our embedded system's restricted computational resources and memory, \networkname is the best architecture to enable on-chip audio super resolution for BCM.

\subsection{Power Consumption}
\label{subsec:PC}
Since the algorithm pipeline can be run in real-time on Arm Cortex-M7, we measured the power consumption of this microcontroller under two circumstances: (1) during audio super resolution and (2) without audio super resolution. We used a power analyzer (EMK850) to measure the average current within one minute. When our super resolution module is active, the average consumption is 498mW (114.4mA at 4.35V). After the audio super resolution module is deactivated, the microcontroller consumes 406mW (93.0mA at 4.37V). Therefore, \networkname, feature extraction and reconstruction consume 92 mW (498mW - 406mW) on average.

\section{Perceptual Speech Quality Evaluation}
\label{sec:study}
In this section, we present the perceptual speech quality evaluation of our method under both quiet and noisy environments. We describe the user study design, participants, and results in this section. Specifically, we conducted two user studies. The first one is to compare the perceived speech audio quality of different machine learning models. The second user study is to evaluate our method's effectiveness against environmental noises. We utilized a with-in subject user study design. We utilized the Friedman test for non-parametric statistical analysis (p < 0.05) and the Wilcoxon signed-rank test for post-hoc analysis (p < 0.05). We utilized the Mann-Whitney U test to evaluate the difference between user groups for statistical analysis (p < 0.05). 

\subsection{Participant}
\label{subsec:participant}
We recruited 20 participants (14 males and 6 females) with an average age of 33.2 (s.d. = 4.8) separated into two groups. The "Golden Ear" (GE) group had 10 participants (6 males and 4 females) with an average age of 34.2 (s.d. = 5.0). They were specialists who were selected and trained to be able to discern subtle differences in audios. The "Non-Golden Ear" (NGE) group had the other 10 amateur listeners (6 males and 4 females) with an average age of 32.1 (s.d. = 4.6). 

The study was conducted in a quiet listening room. During the test, each participant was required to wear headphones (AKG N20 model). A 5-minute break was required after 10 trials. The whole study lasted for 60 minutes. Each participant received a 30 USD gift card for their time and effort. 

\subsection{User Study 1: Speech Audio Quality in Quiet Environment}
\label{subsec:ES}
This user study included 20 trials. Participants listened to an audio clip from the MEMS air-conduction microphone in each trial, which produced the highest speech audio quality. Then they listened to and compared three audio clips, including: 1) original speech audio from the BCM (Original); 2) speech audio processed by the 2D-UNet v1; and 3) speech audio processed by the ATS-UNet. The three audio clips had the same duration, while each set of audio clips lasted between 5 to 10 seconds with an average duration of 8.2 seconds. Then each participant rated the sound quality of these three audio clips by referring to the high-quality audio clip from the MEMS microphone. We utilized a 5-point Likert scale for the rating (5 = very good, 3 = neural, 1 = very bad). The three audio clips were ordered randomly in each trial before the user study. Participants were allowed to listen to and compare audio clips repeatedly. In total, each participant listened to 80 audio clips. 

\subsubsection{Results}
\label{subsec:USR}
\begin{figure}[ht]
  \includegraphics[width=0.9\linewidth]{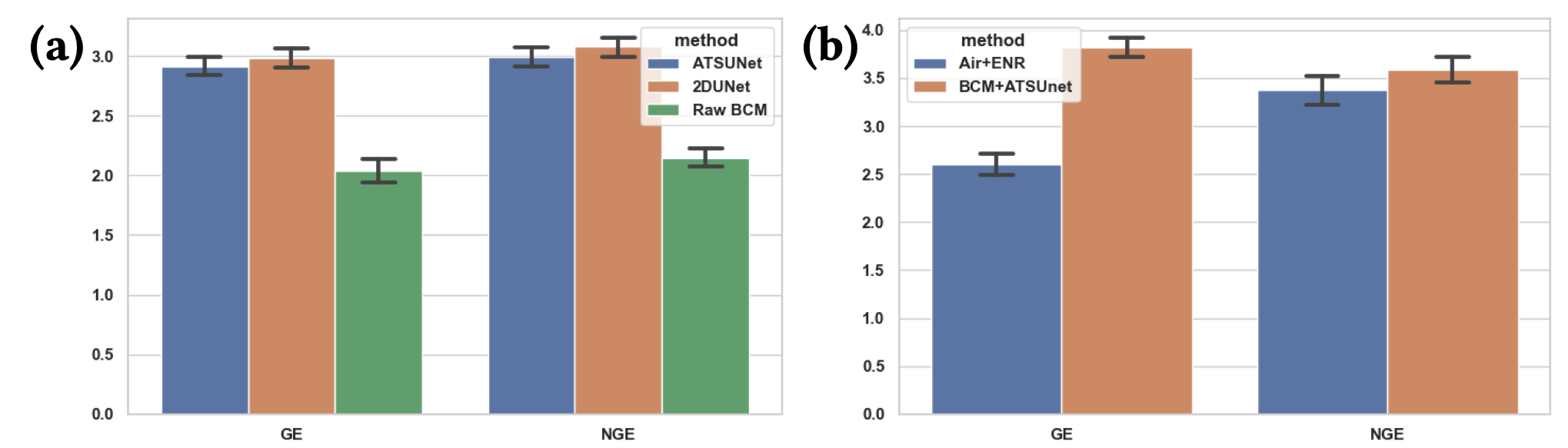}
  \caption{The qualitative speech audio quality evaluation results.}
  \label{fig:evaluation}
\end{figure}

Results show that both 2D-UNet v1 and ATS-UNet can effectively increase the sound quality of audio from the bone conductive microphone. Further, 2D-UNet v1 achieved a better performance than ATS-UNet. As shown in Figure~\ref{fig:evaluation}(a), the mean score of original audio was 2.09 (s.d. = 0.03), of ATS-UNet audio was 2.95 (s.d. = 0.03), and of 2D-UNet v1 audio was 3.03 (s.d. = 0.03). These differences were statistically significant according to a Friedman test ($\chi^2$(2, N = 400) = 376.6, p < 0.001). Post-hoc analysis shows that both the perceived sound quality of audio processed by the ATS-UNet (Z = -13.6, p < 0.001) and the 2D-UNet v1 (Z = -13.9, p < 0.001) significantly outperformed the original bone conductive speech audio. Further, 2D-UNet v1 outperformed ATS-UNet (Z = -2.8, p < 0.01) significantly. 

User group analysis results show that there was significant effect of golden ear status on the perceived sound quality when listening to the original speech audio from the BCM (Z = -2.1, p = 0.036) but not under the 2D-UNet v1 (Z = -0.9, p = 0.37) or ATS-UNet (Z = -1.2, p = 0.23) conditions. 

\subsection{User Study 2:  Effectiveness of \networkname Against Environmental Noises}
\label{subsec:noise}
This user study is to evaluate the effectiveness of our method against environmental noises. We compare our method (BCM+\networkname) to a baseline method --- a single air-conduction microphone with environment noise reduction (AIR+ENR). In this evaluation, we adopted Baseus Encok TWS earbuds - WM01~\footnote{http://www.baseus.com/product-740?lang=en-us} as our comparison baseline. It has a built-in signal processing method for environment noise reduction. During our test with five different brands of earphones with the single speech microphone solution, the Baseus WM01 earbud achieved the best performance in environment noise reduction. 

\subsubsection{Speech Audio Data Collection in Noisy Environments}
We first recruited 5 participants to collect speech audios under various environments, including a noisy pedestrian street, a subway station, and a moving car with the window open. They had an average age of 26.5 (s.d. = 2.5). We used the recording hardware presented in section~\ref{subsec:SADC} to collect the speech audios from both the air-conduction microphone and the bone conduction microphone. Further, we streamed the speech audio from the Baseus WM01 earbud to an iPhone 12 for comparison. We used three hand-clapping events to start each recording and later synchronize the audios. During each data collection session, each participant read the same article that lasts around 2 minutes. The whole data collection procedure lasted 40 minutes. Each participant received a 20 USD gift card. As a result, we collected 2 (minutes) $\times$ 5 (participants) $\times$ 3 (environments) = 30 minutes of speech audios with the bone-conduction microphone, the MEMS air-conduction microphone, and the Baseus WM01 earbud. Among these audio recordings, the raw audios from the MEMS microphone contain stronger environmental noises. 

\subsubsection{Speech Audio Quality Evaluation}
This user study included 24 trials. In each trial, participants listened to an audio clip from the MEMS microphone with stronger background noises. Then they listened to and compared two audio clips, including 1) speech audio from the bone conduction microphone processed by the \networkname, and 2) speech audio from the Baseus WM01 earbud with noise reduction algorithms. Each pair of audio clips had the same duration and lasted between 8 to 13 seconds, with an average duration of 10.96 seconds. Then each participant rated the sound quality of these two audio clips by referring to the audio clip from the MEMS microphone. We utilized a 5-point Likert scale for the rating (5 = very good, 3 = neural, 1 = very bad). The two audio clips were ordered randomly in each trial before the user study. Participants were allowed to listen to and compare audio clips repeatedly. In total, each participant listened to 60 audio clips. 

\subsubsection{Results}
Results show both golden ear (GE) (Z = -12.2, p < 0.001) and non-golden ear (NGE) (Z = -2.2, p =0.02) raters considered the speech audio quality of the BCM+\networkname outperforms the baseline method --- AIR+ENR. As shown in Figure~\ref{fig:evaluation}(b), GE raters scored the speech audio quality of BCM+\networkname with an average of 3.83 (s.d. = 0.86) and of the AIR+ENR with an average of 2.61 (s.d. = 0.93). NGE raters scored the speech audio quality of BCM+\networkname and AIR+ENR with average values of 3.58 (s.d. = 1.17) and 3.38 (s.d. = 1.21) respectively. 

User group analysis results show that there was a significant effect of the user group on the perceived sound quality of AIR+ENR (Z = -7.7, p < 0.001) but not BCM+\networkname (Z = -1.9, p = 0.053). These results indicate that GE and NGE raters perceived similar speech audio quality regarding our method. However, GE raters gave the AIR+ENR a significantly lower score, indicating a poorer preference for AIR+ENR.
\section{Discussion}
\label{sec:discussion}
In this work, we present the first on-chip audio super resolution system for BCM. By integrating a novel ATSM into UNet architecture, \networkname makes it possible to recover the missing high-frequency content captured by the BCM on resource-constrained hearable devices. Therefore, model inferences could be run locally on hearable devices without unwanted data transmission and lower latency. In this section, we discuss potential future works and related applications.

\subsection{Dual Microphone System and Ambient Awareness}
\label{subsec:d1}
Even though BCM is superior to traditional microphones in noisy environments and our system significantly improves the BCM’s audio quality, air conduction microphones still provide higher speech quality in low noise environments. Therefore, a lot of research \cite{liu2004direct, lee2018bone, takada2018self, zhou2020real, yu2020time} has focused on using an air conduction microphone as the primary sensor, accompanied by a BCM for noise reduction. Conversely, low-quality bone conductive audio is used directly in this research, so we hypothesize there may be an opportunity to apply the audio super resolution model on bone conductive speech in conjunction with multi-microphone denoising algorithms. 

BCMs and air conduction microphones are suitable for different scenarios due to their hardware properties. For example, under strong wind noise BCMs are highly desired. Whereas, in a quiet meeting room, BCMs are unnecessary. In this case, the audio super resolution algorithm leads to unnecessary power consumption. Therefore, another potential future research with a dual-microphone system could be ambient awareness. We anticipate a dual-microphone system with ambient awareness can achieve the best user experience with optimal power consumption. With the ambient environment information, we could then determine the appropriate microphone and algorithm combination to be utilized at any instance. 

\subsection{Audio super resolution Applications}
\label{subsec:d2}
In this work, our system incorporated a single BCM, which we model as an audio super resolution problem. We have also observed many other potential real-world applications. For example, recently, many people are wearing masks to prevent COVID-19. While these masks prevent the spread of the virus, it also blocks part of the speech signals. \citet{corey2020acoustic} showed different masks and microphone placements have different impacts on speech quality. We believe the audio super resolution model is a potential solution for recovering the attenuated frequency components from the masks. More and more people pursue high-fidelity music, but for now, the majority of music on the internet is compressed MP3 files. We anticipate our model could be used to recover compression losses generated from lossy compression audio codecs. In general, it is encouraged to try our \networkname if audio quality is degraded by frequency loss.

\subsection{ATSM for other Audio Applications}
\label{subsec:d3}
ATSM was designed for processing spectrograms of the audio signal, one of the most widely used input features for audio-related deep neural networks. Therefore, we believe ATSM can be easily adapt to other audio applications such as speech separation~\cite{Ochiai_2019} and speech emotion recognition~\cite{Drakopoulos_2019}. Researchers can insert ATSM into existing models and reduce the dimension of convolutional layers making the models more lightweight and deployable. Though ATSM was not designed for input features such as waveforms and MFCC, it provides insight on how to enable information flow and enlarge perceptual range without large convolutional kernels.

\subsection{Limitation and Future Work}
\label{subsec:d4}
In this paper, we built a hardware prototype to evaluate the effectiveness of our method to recover high-fidelity speech audio from the bone conduction microphone. The data collection and performance evaluation procedures were performed on the development board, as Figure~\ref{fig:system} shows. We have not developed a wearable hardware solution so that users can wear it comfortably. Further, during our test and evaluation, the placement of the BCM has a significant effect on the audio quality. In our implementation, we utilized an earmuff to stabilize the BCM to the user's skin with its location shown in Figure~\ref{fig:mic}. We chose this location for two reasons. 1) It can pick good quality of speech audios during our pilot study. 2) We referred to the cutting-edge design of modern bone conduction speakers. We expect future work to investigate the optimized location and mounting mechanism. Further, we expect future work to explore sensor fusion approaches to enable better speech audio quality using air and bone conduction microphones.

\section{Conclusion}
\label{sec:conclusion}
In this paper, we proposed a novel real-time embedded audio super resolution based speech capture system with BCM. By integrating a novel ATSM into UNet architecture, \networkname can efficiently process bone conduction speech audio signal with minimal computational resources among our proposed lightweight audio super resolution models. Compared with the baseline method, \networkname achieved higher performance in audio quality and reduced the size by approximately 100 times. With the reduction in computation complexity, our system can achieve real-time processing on a Cortex-M7 with an average power consumption of 92 mW. User studies demonstrate that our system significantly improves the perceptual quality of bone conductive speech. We believe our system will promote the usage of BCM in earphones and other deep learning-based audio processing applications, particularly those deployed on resource-constrained embedded systems.

\bibliographystyle{ACM-Reference-Format}
\bibliography{ref}

\end{document}